# REGARDING THE POTENTIAL IMPACT OF DOUBLE STAR OBSERVATIONS ON CONCEPTIONS OF THE UNIVERSE OF STARS IN THE EARLY 17TH CENTURY

Christopher M. Graney
Jefferson Community & Technical College
Louisville, Kentucky 40272 (USA)
christopher.graney@kctcs.edu
www.jefferson.kctcs.edu/faculty/graney

Henry Sipes
Samtec Inc.
540 Park East Blvd.
New Albany, IN 47150 (USA)
henry.sipes@samtec.com

ABSTRACT

Galileo Galilei believed that stars were distant suns whose sizes, measured via his telescope, were a direct indication of distance -- fainter stars (appearing smaller in the telescope) were farther away than brighter ones. Galileo argued in his *Dialogue* that telescopic observation of a chance alignment of a faint (distant) and bright (closer) star would reveal annual parallax, if such double stars could be found. This would provide support both for Galileo's ideas concerning the nature of stars and for the motion of the Earth. However, Galileo actually made observations of such double stars, well before publication of the *Dialogue*. We show that the results of these observations, and the likely results of observations of any double star that was a viable subject for Galileo's telescope, would undermine Galileo's ideas, not support them. We argue that such observations would lead either to the more correct conclusion that stars were sun-like bodies of varying sizes which could be physically grouped, or to the less correct conclusion that stars are not sun-like bodies, and even to the idea that the Earth did not move. Lastly, we contrast these conclusions to those reached through applying Galileo's ideas to observations of visible stars as a whole.

## 1. INTRODUCTION

*The stars are suns at large and varying distances from Earth, and it should be possible to obtain data supporting this view from observing a double star, should any double stars be found.* These are among the arguments Galileo Galilei makes in his 1632 *Dialogue Concerning the Two Chief World Systems -- Ptolemaic & Copernican*. However, Galileo had observed double stars as early as 1617, and those observations could not have yielded the sort of data he describes in the *Dialogue*. In this paper we shall review Galileo's double star observations, we shall investigate what additional observations of this kind by Galileo would have revealed, and we shall close with some discussion of the impact that an early thorough telescopic study of stars might have had on conceptions regarding stars and the structure of the universe.

## 2. GALILEO ON THE STARS

The modern idea that the stars are the same class of object as the sun, just much further distant, emerged shortly after Copernicus introduced his heliocentric theory. Galileo, speaking through his character of Salviatti, states this idea in the *Dialogue*:

> See, then, how neatly the precipitous motion of each twenty-four hours is taken away from the universe, and how the fixed

> stars (which are so many suns) agree with our sun in enjoying perpetual rest [p. 327].

If the stars are all suns, then differences in the brightness of the stars must be due to differences in distances, with fainter stars being further away.  This view of the universe's structure was not part of Copernicus' original theory -- in *De Revolutionibus Orbium Coelestium* (1543) Copernicus places a sphere of stars beyond the orbit of Saturn; the decades following Copernicus saw the introduction of the notion that the stars were more than simply part of a spherical “firmament” but rather were suns scattered through space (Figure 1).  This idea made the stars evidence of a universe with a vast and perhaps limitless structure beyond the solar system.

Galileo argues in the *Dialogue* that this idea could be tested by telescopic observation of certain double stars (ones in which the two component stars differ significantly in magnitude), and that such a test would also demonstrate that the Earth was moving.  He compares the annual parallax of stars to the retrograde motions of the outer planets, and then argues that such parallax would be most easily detectable in two stars that lay along the same line of sight but at different distances:

> For I do not believe that the stars are spread over a spherical surface at equal distances from one center; I suppose their distances from us vary so much that some are two or three times as remote as others. Thus if some tiny star were found by the telescope quite close to some of the larger

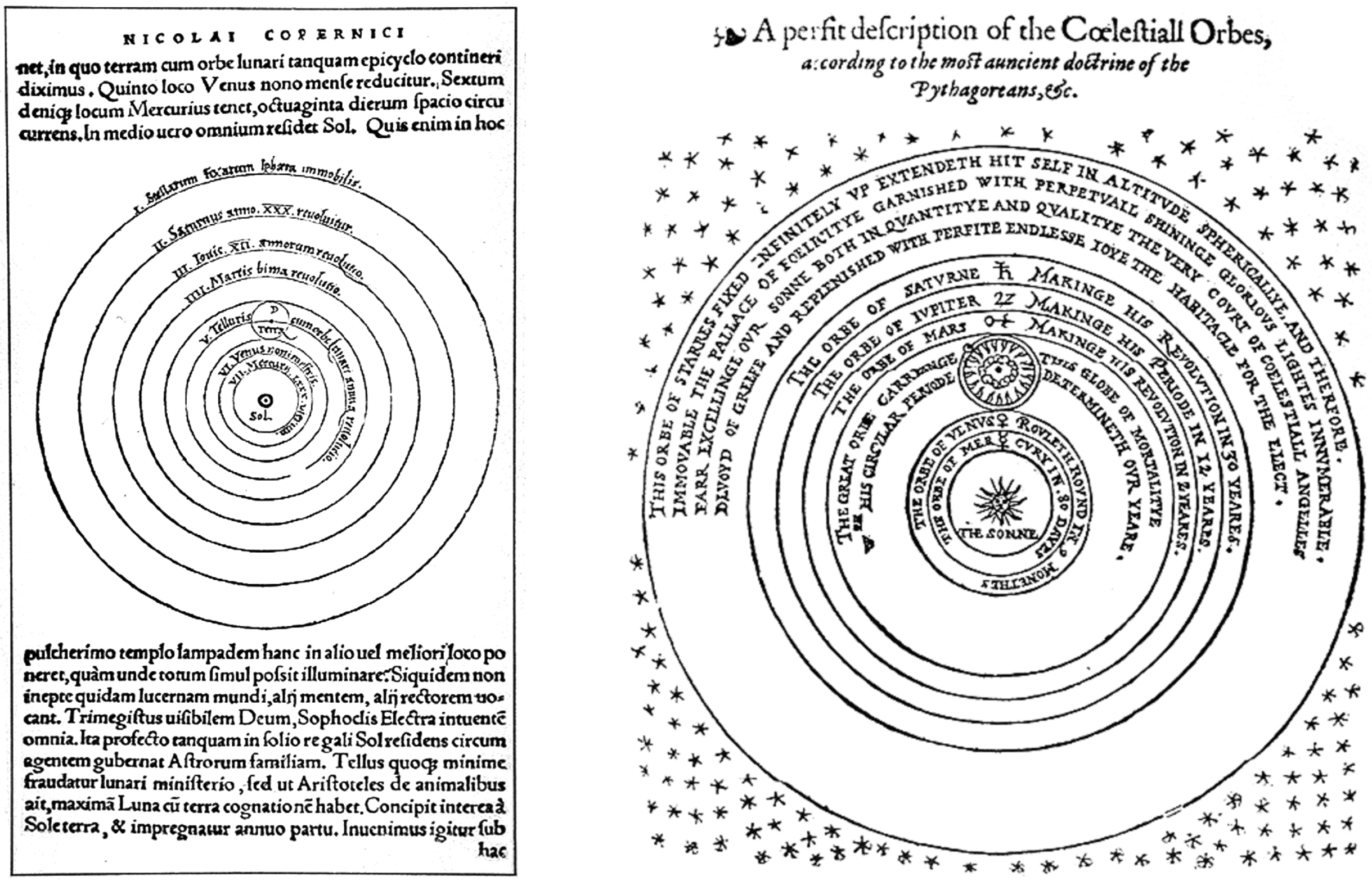

**FIGURE 1: Left -- Illustration from Copernicus' *De Revolutionibus Orbium Coelestium* (1543), showing the stars as a sphere (a firmament) beyond the orbit of Saturn. Right -- Illustration of the Copernican system from Thomas Digges' *A Perfit Description of the Coelestiall Orbes* (1576) showing the Copernican system with the stars extending outward indefinitely. Giordano Bruno in his 1584 *La Cena De Le Ceneri* described the stars as Digges envisioned, but widely separated.**

> ones, and if that one were therefore very remote, it might happen that some sensible alterations would take place among them corresponding to those of the outer planets [i.e. parallax; pp. 382-383].[1]

1 The reader should note that the term “double star” when used with Galileo's observations refers to double stars as envisioned by Galileo -- two stars at significantly differing distances along the same line of sight. While modern astronomers usually take “double star” to mean stars close in all three dimensions of space, the method Galileo proposed in the *Dialogue* of using two stars along a line of sight to detect parallax is commonly referred to as “Galileo's double star method” and However, what Galileo does not reveal in the *Dialogue* is that such an observational test had already been made, by Galileo, with negative results.

Galileo's telescopic observations of stars in regards to the parallax question have recently been discussed by Ondra (2004) and Siebert (2005); the high quality of Galileo's observations has been discussed by Standish and Nobili (1997) and Graney (2007); the high optical quality of Galileo's telescopes has been discussed by Greco, Molesini, and Quercioli (1992). To briefly summarize these discussions: Galileo is known to have observed a number of multiple-star systems. These include very precisely recorded observations in 1617 of the double star Mizar in Ursa Major and of the Trapezium in Orion. In the case of Mizar Galileo measured the component stars to have apparent angular diameters of 6” and 4”, and a separation of 15”; in the case of the Trapezium he made a precise sketch of the region, accurately mapping the positions of stars separated by as little as 15”, and noting relative sizes (without giving absolute size measurements). Both the Mizar measurements and the Trapezium sketch are in excellent agreement with modern data on these stars' positions. In the case of Mizar, the diameters Galileo recorded are consistent with the diameters of Airy disks that would be formed by telescopes of the size he used. It appears that Galileo could measure positions and sizes to an accuracy of 2”.

---

so that is how the term “double star” will be used in this paper.

It is clear that Galileo drew very specific conclusions from his observations and measurements of stars. Galileo interpreted the images of stars seen through his telescopes as being the physical globes of stars. From that interpretation it necessarily follows that stellar sizes measured with the telescope are inversely related to stellar distances. Galileo assumed that the size of a star was approximately equal to the size of the sun. Using this method he believed he could determine the distance to any star whose apparent angular diameter he could measure: $L = \alpha_{\odot}/\alpha$, where $L$ is the distance to the star in AU, $\alpha$ is the angular diameter of the star as measured via his telescope, and $\alpha_{\odot}$ is the angular diameter of the sun.

These ideas may surprise the reader -- modern discussions of Galileo's major astronomical work tend either to overlook Galileo's views and discoveries concerning the stars, or to limit themselves to his description of them in the *Starry Messenger* (1610), his first publication of his telescopic observations.[2] In the *Starry Messenger* Galileo describes his early impression of stars as seen through the telescope:

> The fixed stars are never seen to be bounded by a circular periphery, but rather have the aspect of blazes whose rays vibrate around them and scintillate a great deal. Viewed with the telescope they appear of a shape similar to that which they

2 As an example, the *Cambridge Companion to Galileo* contains a number of articles, including “Galileo's discoveries with the telescope and their evidence for the Copernican theory.” These discuss in detail Galileo's observations of the moon, Jupiter's satellites, Venus' phases, and sunspots, but mention the stars only briefly, and only in terms of the *Starry Messenger*.

> present to the naked eye, but sufficiently enlarged so that a star of fifth or sixth magnitude seems to equal the Dog Star, largest of all the fixed stars [p. 46].

However, this impression quickly fell away, and by 1617 Galileo was thinking of stars in terms of $L = \alpha_{\odot}/\alpha$. His observing notes on Mizar contain the following measurements -- separation: 0°, 0', 15”; larger star radius: 0°, 0', 3”; smaller star radius -- 2”; gap between them -- 10”. He also notes that the radius of the sun contains 300 radii of the larger star, so therefore the distance to the star contains 300 distances to sun, if the star is the size of the sun (*Opere*, III, p. 877).[3] This view of the stars appears again in Galileo's 1624 “Reply to Ingoli”:

> I say that if you measure Jupiter's diameter exactly, it barely comes to 40 seconds, so that the sun's diameter becomes 50 times greater; but Jupiter's diameter is no less than ten times larger than that of an average fixed star (as a good telescope will show us), so that the sun's diameter is five hundred times that of an average fixed star; from this it immediately follows that the distance to the stellar region is five hundred times greater than that between us and the sun [p. 167]....

---

3 The text of Galileo's notes:

> *Inter mediam caudae Elicis et sibi proximam pono nunc gr. 0.0'.15”.*
> *Semidiameter stellae maioris gr. 0.0.3", minoris vero 2", et intercapedo 10"....*
> *Semidiameter ⊙ continet semidiametros stellae maioris 300.*
> *Distantia ergo stellae continet distantias ⊙ 300 (si stella ponatur tam magna ut ⊙)....*

Elicis is Helice -- the Great Bear.

> [M]any years ago that I learned by sensory experience that no fixed star subtends even 5 seconds, many not even 4, and innumerable others not even 2 [p. 174]....
>
> I do not think the fixed stars are are all placed on a spherical surface, so as to be equidistant from a particular point, such as the center of a sphere; indeed only God knows whether for any group larger than three there is a single point from which they are equidistant [p. 176]....

Finally, these ideas appear in the *Dialogue* of 1632:

> [T]he apparent diameter of the sun at its average distance is about one-half a degree, or 30 minutes; this is 1,800 seconds, or 108,000 third-order divisions. And since the apparent diameter of a fixed star of the first magnitude is no more than 5 seconds, or 300 thirds, and the diameter of one of the sixth magnitude measures 50 thirds..., then the diameter of the sun contains the diameter of a fixed star of the sixth magnitude 2,160 times. Therefore if one assumes that a fixed star of the sixth magnitude is really equal to the sun and not larger, this amounts to saying that if the sun moved away until its diameter looked to be 1/2160th of what it now appears to be, its distance would have to be 2,160 times what it is in fact now. This is the same as to say that the distance of a fixed star of sixth magnitude is 2,160 radii of the earth's orbit [p. 359-360].

## 3. PARALLAX PARADOX

But there is an inherent contradiction between Galileo's assumptions and his data. Distances of hundreds or thousands of AU doubtlessly seemed far in the early 17$^{th}$ century, and Galileo argued that such distances supported Copernican ideas[4], but in fact any star at such a distance would have significant annual parallax, and any two stars lying at substantially different such distances along the same line of sight would exhibit annual variations in their separations easily detectable to an observer capable of measuring to 2” accuracy.

For example, let us look at Galileo's Mizar data. Operating on the assumption that both stars are identical to the sun, if Mizar A, with an angular diameter of $\alpha_A$ = *6”*, is 300 AU distant as Galileo calculated, then Mizar B, at $\alpha_B$ = *4”*, is 450 AU distant. Angles $\phi_A$ and $\phi_B$ (Figure 2) are then 688” and 458”, respectively, with the difference between them ($\Delta = \phi_A - \phi_B$) being 230”. As this is two orders of magnitude greater than what Galileo can measure, the effects of annual parallax should be almost immediately observable -- the two stars should swing around each other dramatically.

As Mizar shows no such annual motions, what conclusions can be drawn? Answering this parallax paradox, while retaining Galileo's general conceptions regarding the structure of the universe beyond the solar system, requires challenging the assumptions Galileo made about the sizes of stars.

One route out of the paradox is to invoke the time-honored claim that parallax is not seen because the stars are too distant. This was how Copernicus explained the absence of observable parallax in a heliocentric universe:

---

4 For example, see *Dialogue*, p. 359 as well as “Reply to Ingoli”, pp. 166-168.

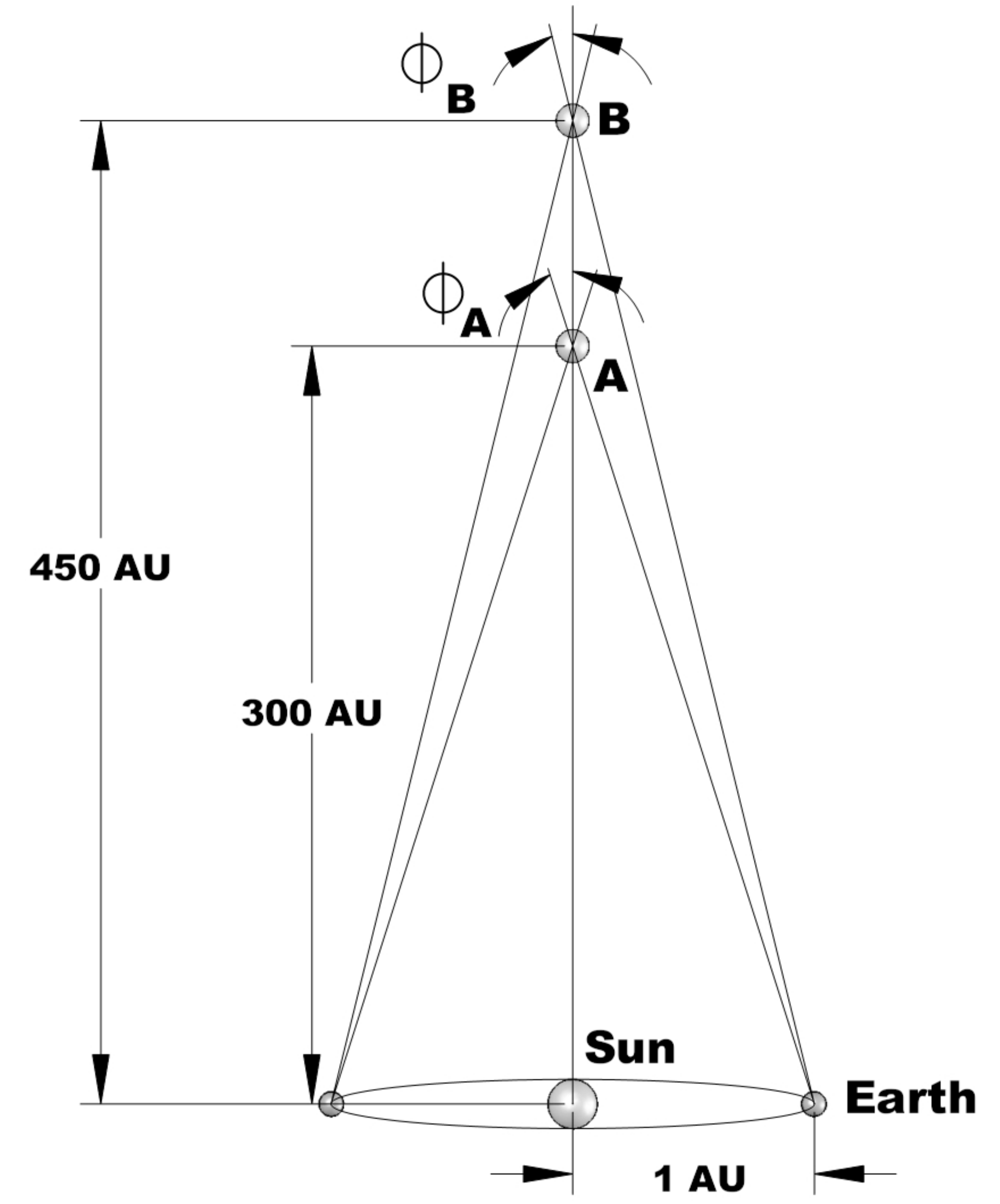

**FIGURE 2: Parallax in Mizar, according to Galileo's ideas.**

> But that there are no such appearances [parallax] among the fixed stars argues that they are at an immense height away, which makes the circle of annual movement [of the Earth] or its image disappear from before our eyes since every visible thing has a certain distance beyond which it is no longer seen, as is shown in optics. For the brilliance of their lights shows that there is a very great distance between Saturn the highest of the planets and the sphere of the fixed stars. It is by this mark in particular that they are distinguished from the planets, as it is proper to have the greatest difference between the moved and the unmoved [*On the Revolutions of the Heavenly Spheres*, p. 27].

Indeed, if Mizar A and B lay a hundred times farther away than Galileo calculated, at 30,000 AU and 45,000 AU respectively, then $\Delta \approx 2''$ and we could understand how parallax might escape easy detection by an observer who can measure to 2" accuracy. But under Galileo's interpretation of the images of stars seen through his telescopes being the physical bodies of those stars, this hundredfold increase in distance would require a hundredfold increase in stars' physical sizes in order to explain the stellar diameters measured by Galileo. Thus this route out of the paradox leads to the "ad hoc" creation of a new class of giant celestial objects, with the sun and stars having little in common. It is also completely at odds with Galileo's ideas about the stars being suns.

A second route out of the paradox is to retain the idea of sun-like stars but to allow for some variation in their sizes. If the measured difference in diameters of Mizar A and Mizar B is due to the two stars differing in real size, the lack of parallax can be explained by the two stars being paired and therefore not lying at greatly varying distances. Assuming Mizar A is equal in size to the sun ($d_A = 1\ d_\odot$) and therefore at $L = 300\ AU$ distant, for $\Delta \leq 2''$ Mizar B must lie at 300 AU to within $s_1 = 0.875\ AU$ (Figure 3), and be 2/3 solar diameters in size ($d_B = 2/3\ d_\odot$). The 15" angular separation of the two stars would translate into $s_2 = 2.5\ d_\odot$ (Figure 3); any significant separation between the two component stars would have to fall along the line of sight. On the other hand, if Mizar B is assumed to be equal in size to the sun ($d_B = 1\ d_\odot$, $L = 450\ AU$), then $\Delta \leq 2''$ requires $d_A = 1.5\ d_\odot$, $s_1 = 1.98\ AU$, $s_2 = 3.75\ d_\odot$. Note that Galileo's assumptions are less challenged by taking the brighter star to be the size of the sun. The more variation possible in the sizes of stars, the more Galileo's entire method of determining stellar distance is undermined.

## 4. A MATHEMATICAL MODEL FOR “ADDITIONAL” GALILEAN DOUBLE-STAR DATA

The second route out of the parallax paradox seems the preferable one insofar as it leaves Galileo's ideas about stars being suns relatively intact. Nonetheless, perhaps we are basing too much on one double star; might other stars yield other results? Unfortunately, the 1617 Mizar observation is the only known instance where Galileo recorded precise measurements of both diameters and separations of a double or multiple star system (the other known observations involve either simply a recording of positions or a simple description -- Siebert 2005).

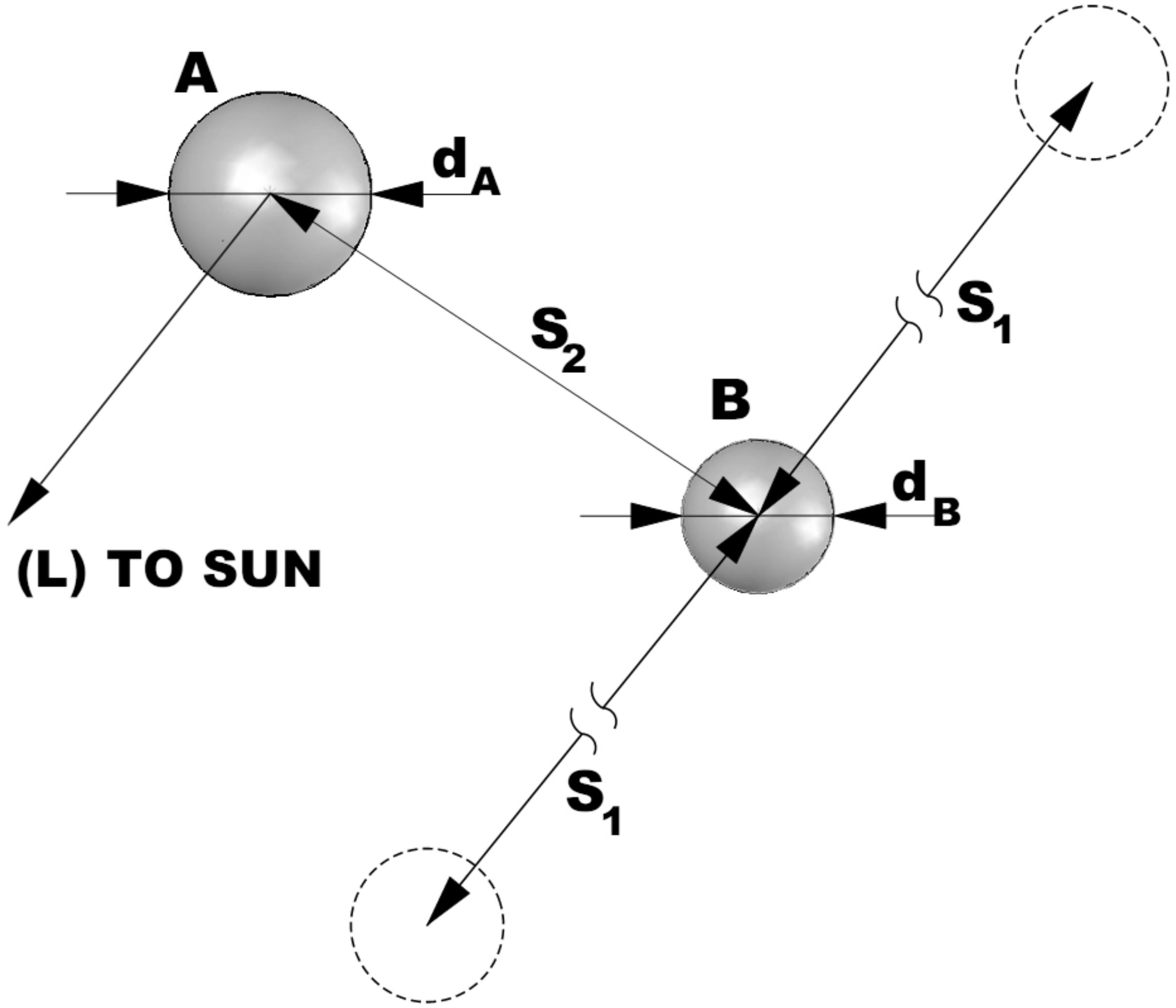

**FIGURE 3**

Nevertheless, the 1617 Mizar observation can help us gain insight into what other observations might have revealed had Galileo undertaken a systematic search for double stars. We can use the Mizar data to create a mathematical model of the telescope, eye, and sky system Galileo used to observe Mizar, and then use that model to "observe" other double stars that would have been turned up by a thorough search for such stars using a Galilean telescope -- the kind of search that might have been undertaken in a determined effort to detect parallax in double stars. With this model we can gain insight into what a 1617 investigation of double stars would have yielded in the way of results.

Galileo built telescopes that were "optically perfect" (Greco, Molesini, Quercioli 1992), so we base our model on the assumption that star images formed by the telescope Galileo used to observe Mizar in 1617 are textbook diffraction patterns for a circular aperture, whose intensity varies as $I(r) = I_0[J_1(r)/r]^2$ where $J_1(r)$ is a Bessel function of the first kind. $I(r)$ falls to zero at the Airy disk radius $r_A = 1.22\lambda/D$ . Here $D$ is the telescope's aperture, and $\lambda$ is the wavelength of light (which in this paper we will take as 550 nm, the center of the visible spectrum). The visible Airy disk will be smaller than $r_A$ due to the intensity falling below the threshold of detection of the eye. So while Mizar A and B, along with every star viewed through the same telescope, share the same $r_A$, the visible Airy disk of Mizar A will be larger than that of Mizar B (Figure 4).

To form a model of the telescope-eye-and-sky system Galileo used to observe Mizar (all three affected how Mizar appeared to Galileo) in 1617, we used the magnitudes of Mizar A and Mizar B to produce relative intensity curves, and then fit a detection threshold $t$ and aperture $D$ so as to reproduce Galileo's Mizar measurements of $\alpha_A = 6''$, $\alpha_B = 4''$ (Figure 4). The magnitudes of any double star can then be fed into the model to produce an estimate of the $\alpha_A$ and $\alpha_B$ measurement Galileo would have obtained from observing that star using the same telescope-eye-sky system

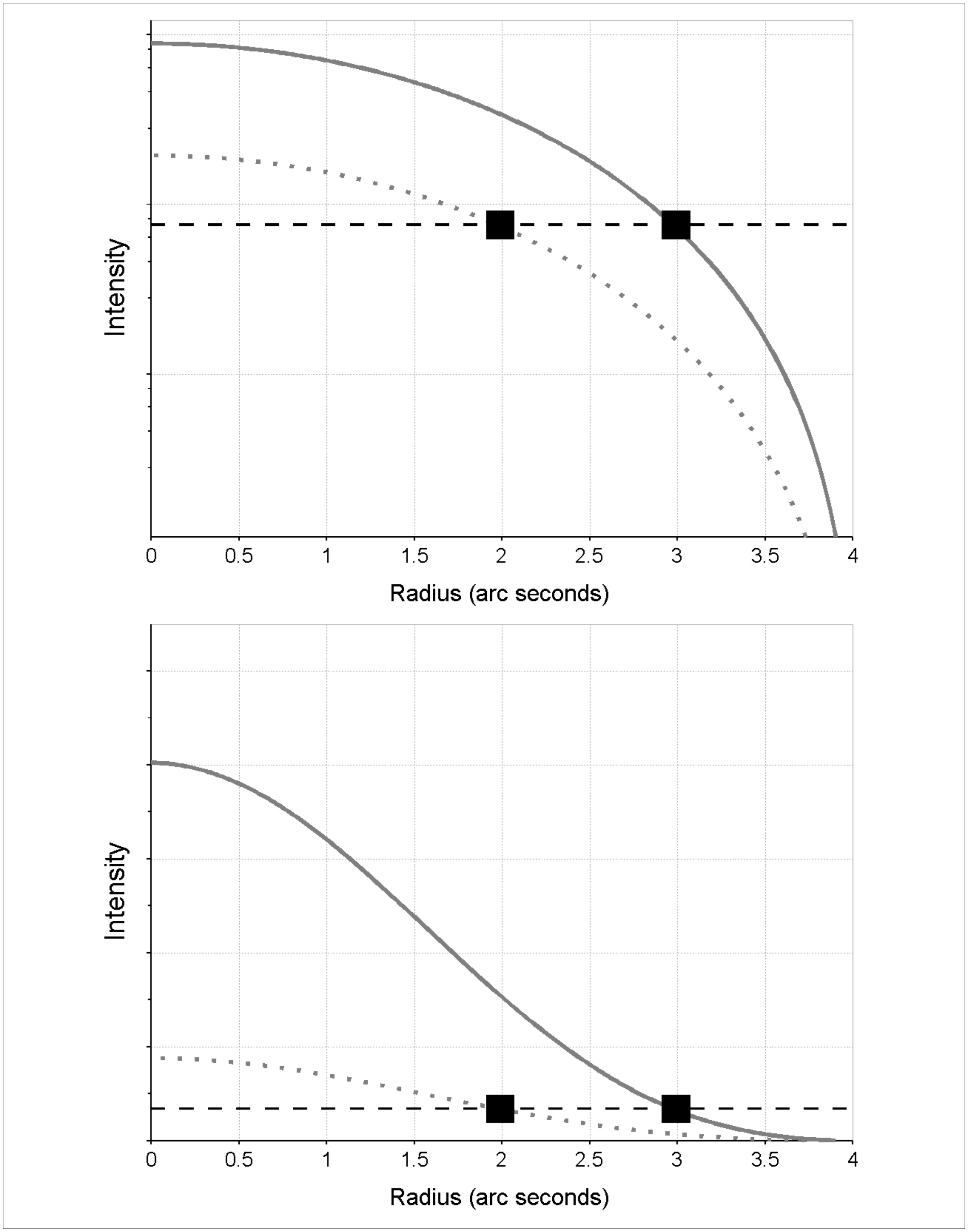

**FIGURE 4 (previous page): The plots show curves for diffraction pattern relative intensity vs. radius for Mizar A (solid) and Mizar B (dotted) using magnitudes from the *Washington Double Star Catalog (WDS)*. The upper plot is on a log axis, the lower on a linear axis. The horizontal dashed line is detection threshold *t*, chosen to yield an $\alpha_A/\alpha_B$ ratio of 3/2 for Mizar, matching Galileo's measurements. We chose a circular aperture *D* for the model to yield Galileo's Mizar image sizes for $\lambda$ = *550 nm*. Results are *D = 34.0 mm*, *t = mag 4.9*. Readers may feel, as we did, that our model's D value compares reasonably to Galileo's actual telescopes (Greco, Molesini, Quercioli 1992), but that *t* seems wrong. After all, unaided eyes can detect stars of sixth magnitude, and clearly even a small telescope will reveal much fainter stars still. Nonetheless, the physics of diffraction through a circular aperture is well-established, and the fit of *t* is entirely a factor of $\alpha_A/\alpha_B$; it is independent of $\lambda$ or *D*. Thus *t* must be explained in terms of the eye or sky conditions at the time the Mizar observation was made. We might explain *t* by supposing that Galileo made his Mizar measurements during twilight or under poor skies. However, as is explained in Figure 6, when we tested the model by making visual $\alpha_A/\alpha_B$ measurements under clear dark skies our results generally supported the model, ruling out poor skies. As discussed in Figure 6, a possible explanation is that the eye is known to have two different sorts of detection cells (commonly known as "rods" and "cones"), and that our *t* value represents a threshold for the less sensitive of the two. At any rate, the *t* value yielded reasonable results.**

he used to observe Mizar. From that information and the star's separation we can calculate the same information we obtained for Mizar: $L$, $d_B$, $s_1$, and $s_2$. We can also construct a simulation of what Galileo would see through the eyepiece, and a diagram of the double star system calculated using Galileo's methods.

Figure 5 and Table 1 show the results of running magnitudes and separations of selected double stars (data from the *Washington Double Star Catalog -- WDS*) through the model. We selected stars from the *WDS* to “observe” with the model based on several criteria. First, the stars had to be bright, under the assumption that any search for double stars would begin with bright stars simply as an issue of number. Our particular criterion was that the brighter component of the double be magnitude 4.0 or brighter. This puts the number of candidate stars Galileo would have to examine in a search well under 1000 (Hoffleit 1991)-- by comparison, William Herschel's parallax-motivated search for double stars yielded almost 1000 actual double stars, many of which were quite faint (Hirshfeld 2001, pp.179-188). Second, the two stars had to be seen as separate stars in Galileo's telescope. We chose stars whose separations in 1617 would be 4.0” or greater. Acknowledging the field of view of Galileo's telescopes (Drake and Kowal 1980, p. 77), the separations had to be smaller than 400”. Third, in order to work with the model the dimmer component of the double had to be brighter than magnitude 4.9 (refer to the caption of Figure 4). Last, in order for the stars to have differing apparent diameters they have to have differing magnitudes; we chose stars whose components differed by at least half a magnitude.

Readers may question our use of some widely separated pairs, and whether these would be valuable to Galileo in a parallax search; could Galileo note changes in position between such widely-space stars? We gave Galileo the benefit of the doubt (Galileo did accurately record positions of Jupiter's

# β Cyg

WDS DATA

| WDS Number | $V_A$ | $\Delta m_V$ | $V_B$ | separation |
|---|---|---|---|---|
| 19307+2758 | 3.19 | 1.49 | 4.68 | 34.4 |
| | magnitudes | | | *in seconds* |

DATA FROM MODEL

| $\alpha_A$ | $\alpha_B$ | $\alpha_A/\alpha_B$ | L | $d_B$ | $s_1$ | $s_2$ | $s_2$ |
|---|---|---|---|---|---|---|---|
| 5.0 | 1.9 | 2.6 | 360 | 3/8 | 1.261 | 6.88 | 0.031 |
| *in seconds* | | | *in AU* | *in $d_\odot$* | *in AU* | *in $d_\odot$* | *in AU* |

VISUAL DATA

| $\alpha_A/\alpha_B$ |
|---|
| 2.0 |

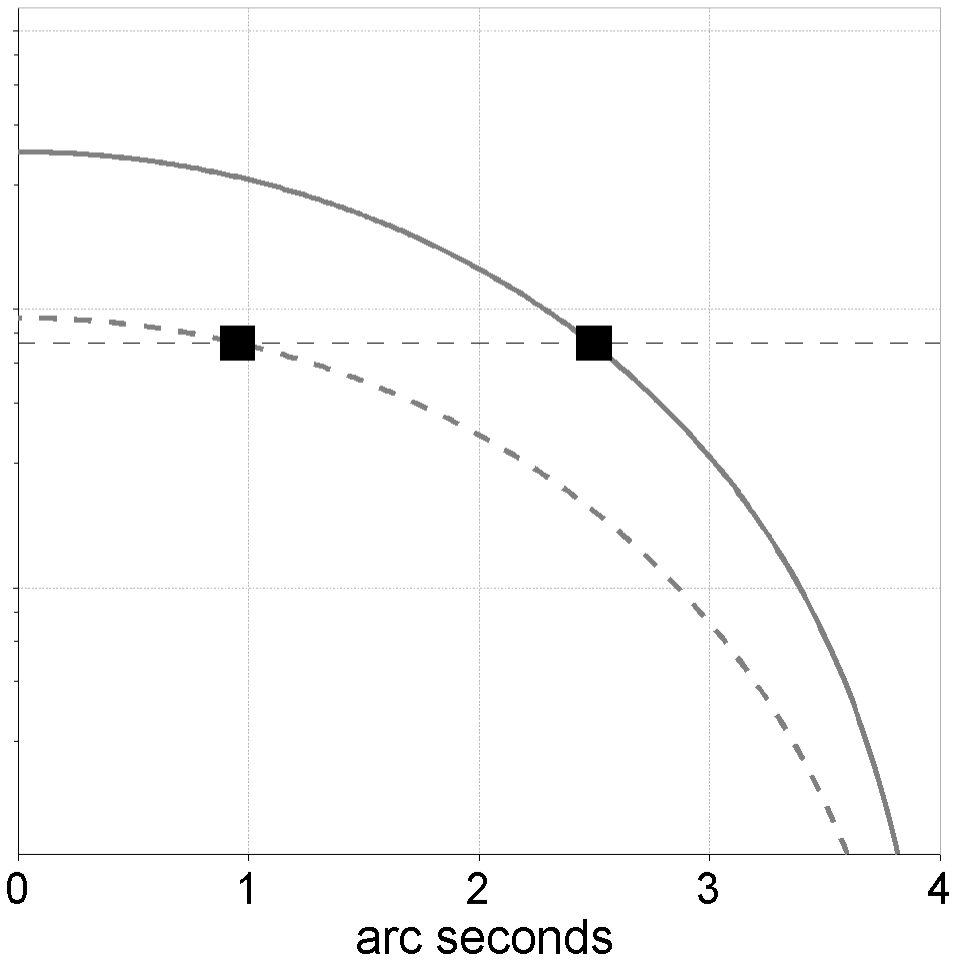

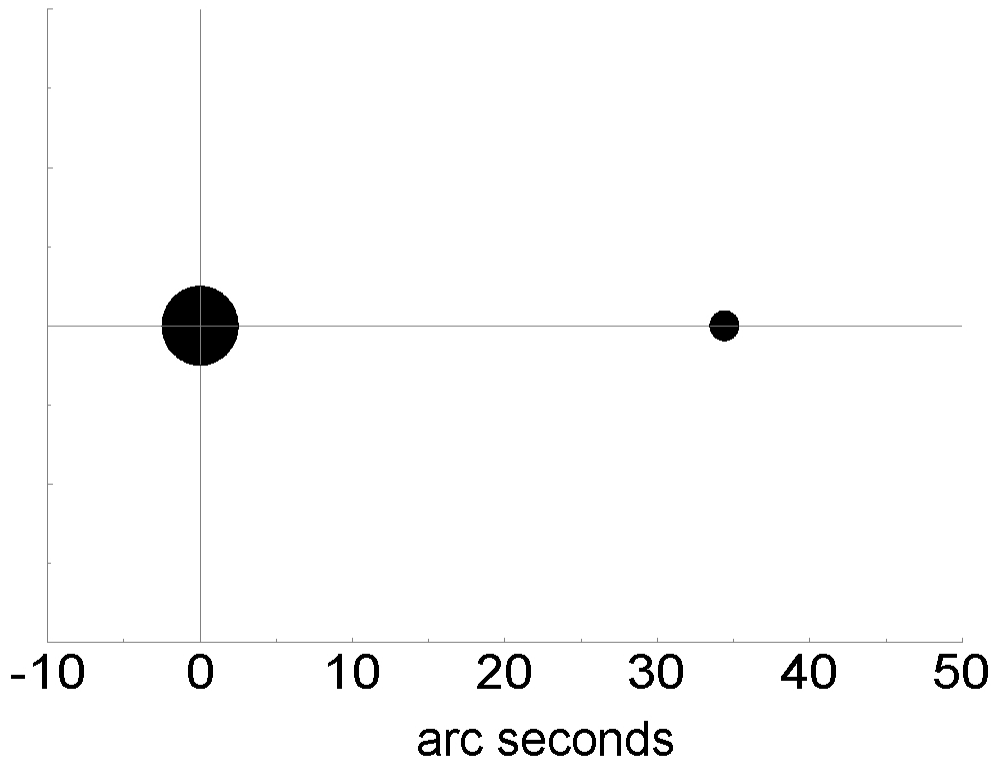

**FIGURE 5: Results of "observing" the double star ß Cyg with the mathematical model are presented here. Here we have:**

1. ***WDS* data on magnitudes and separation.**
2. **$\alpha_A$, $\alpha_B$, and $\alpha_A/\alpha_B$ calculated via the model, as well as distance *L*, size of the smaller companion $d_B$, and separations $s_1$ and $s_2$ calculated using Galilean methods as discussed in section 3 of this paper [calculated for the case of $d_A = 1\ d_\odot$, which, as mentioned in section 3, results in the least deviation from Galileo's assumptions; conversions between $d_\odot$ and AU were done using Galileo's value of *1 AU = 110 $d_\odot$* (*Dialogue*, p. 360)].**
3. **Diffraction pattern relative intensity curves used to calculate $\alpha_A$ and $\alpha_B$ (log plot).**
4. **Our visual measurement of $\alpha_A/\alpha_B$ used to corroborate the results of the model.**
5. **A simulated telescopic view of the star based on the model's output.**

Data from the Washington Double Star Catalog

| Star | WDS Number | magnitudes | | | in seconds |
|---|---|---|---|---|---|
| | | $V_A$ | $\Delta m_V$ | $V_B$ | separation |
| θ Tau | 04287+1552 | 3.41 | 0.53 | 3.94 | 336.7 |
| α Gem‡ | 07346+3153 | 1.93 | 1.04 | 2.97 | 6.9 |
| ζ U Ma$ | 13239+5456 | 2.23 | 1.65 | 3.88 | 14.3 |
| β Sco | 16054-1948 | 2.59 | 1.93 | 4.52 | 13.1 |
| β Cyg | 19307+2758 | 3.19 | 1.49 | 4.68 | 34.4 |
| o Cyg | 20136+4644 | 3.93 | 0.90 | 4.83 | 336.4 |
| α Cap | 20181-1233 | 3.67 | 0.67 | 4.34 | 381.2 |

Data calculated from model or obtained visually

| Star | in seconds | | calculated | visual | in AU | in $d_\odot$ | in AU | in $d_\odot$ | in AU |
|---|---|---|---|---|---|---|---|---|---|
| | $\alpha_A$ | $\alpha_B$ | $\alpha_A/\alpha_B$ | $\alpha_A/\alpha_B$ | L | $d_B$ | $s_1$ | $s_2$ | $s_2$ |
| θ Tau | 4.7 | 3.9 | 1.2 | 1.1 | 383 | 5/6 | 1.427 | 71.64 | 0.651 |
| α Gem‡ | 6.2 | 5.2 | 1.2 | 1.3 | 290 | 5/6 | 0.818 | 1.11 | 0.010 |
| ζ U Ma$ | 6 | 4 | 1.5 | 1.3 | 300 | 2/3 | 0.875 | 2.38 | 0.022 |
| β Sco | 5.6 | 2.5 | 2.2 | ~ | 321 | 4/9 | 1.002 | 2.34 | 0.021 |
| β Cyg | 5.0 | 1.9 | 2.6 | 2.0 | 360 | 3/8 | 1.261 | 6.88 | 0.063 |
| o Cyg | 3.9 | 1.1 | 3.5 | 1.5 | 462 | 2/7 | 2.079 | 86.26 | 0.784 |
| α Cap | 4.3 | 3.0 | 1.4 | 1.3 | 419 | 5/7 | 1.709 | 88.65 | 0.806 |

$Mizar -- Used to fit parameters in model

‡Separation values calculated for the year 1617

~ β Sco was not observable during the time we worked on this project moons over similar scales -- Standish and Nobili 1997). Readers may also wonder why we did not look at other close pairings such as can be found in the Pleiades star cluster. In selecting double stars we limited ourselves to *WDS* data, which, when we applied our criteria, yielded the candidate stars listed in Table 1. We chose not to hunt for other candidate stars, although a check of other sources indicates the *WDS* data has yielded most of the likely candidates.

**TABLE 1: Results of mathematical model calculations for all the stars selected from the *WDS*. All work done as in Figure 5.**

We made visual observations to corroborate the results of the model (refer to Figure 6). Our initial goal in making the observations was merely to make a rough check of the model's results. However, the visual observations gave us added confidence in the results of the model, added appreciation for Galileo's abilities at the eyepiece, and added understanding of why he interpreted the images he saw to be the physical globes of stars.

## 5. POSSIBLE CONCLUSION: DOUBLE STAR SYSTEMS EXIST

The selected double stars and their characteristics as determined using the model and Galileo's methods show that additional observations of double stars by Galileo would have been likely to yield results consistent with the Mizar data. Note that in general $s_1$ is considerably less than $s_2$ for all stars, suggesting to Galileo that it is unlikely that significant line-of-sight separations exist between the stars as that would imply a preference to align toward Earth. Galileo

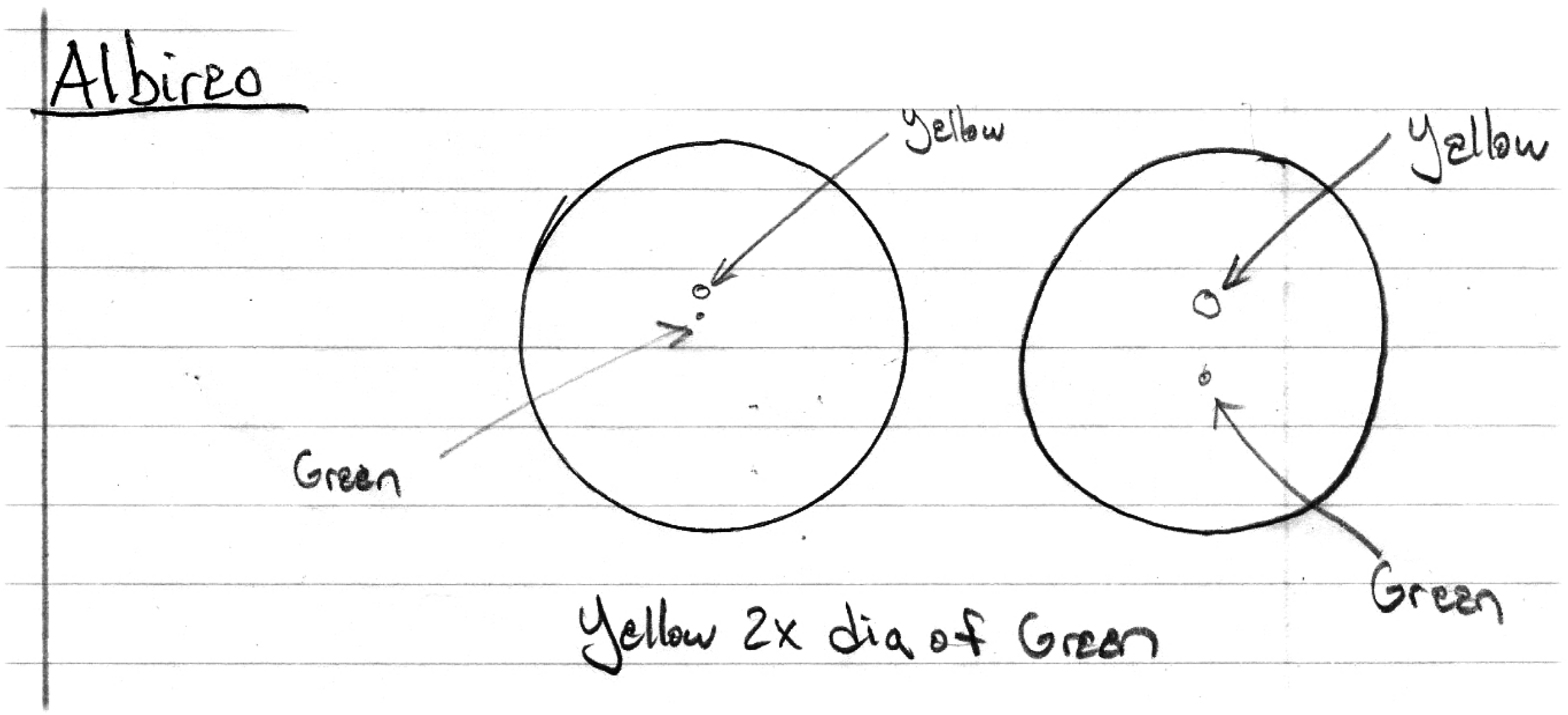

**FIGURE 6:** We attempted to test the model's results using visual observations with a 100 mm f/15 refractor outfitted with a 35 mm mask. Our initial goal was only to attempt a rough confirmation of the model's results; we did not hope to be able to match Galileo's skill and our intent was not to replicate his work. Author Sipes made the observations, recording the ratio of apparent sizes of component stars ($\alpha_A/\alpha_B$). This figure shows simple notes from Sipes's visual observations (left, 24 mm eyepiece, right 8 mm). Of particular interest was the disk-like appearance of stars as seen through the telescope when masked to a 35 mm aperture. Sipes felt the appearance was as clearly disk-like as the image of a planet, except in the few cases when a diffraction ring was noticeable. The star image disks definitely responded to magnification, as is noticeable even in these rough sketches.

Sipes' visual estimates of $\alpha_A/\alpha_B$, which he made independently and without prior knowledge of the results of the model, compared favorably with the $\alpha_A/\alpha_B$ yielded by the model (independently calculated by author Graney). This was true even though the star pairs observed varied considerably in both brightness and separation (see θ Tau, α Gem, and α Cap in Table 1). The visual observations also shed light on the question of the detection threshold level $t$ (see Figure 4). The observations were all made under **clear dark skies, meaning that the *t = mag 4.9* value is not due to poor or bright skies as mentioned in Figure 4. As mentioned in Figure 4, a possible explanation for the *t* threshold is that the eye is known to have two different sorts of detection cells (commonly known as "rods" and "cones"), and that our *t* value represents a threshold for the less sensitive of the two. During our observations, brighter stars appeared as definite disks when viewed through the telescope and 35 mm mask, even at low magnifications; they also showed color, and the disk diameters obviously varied with magnitude. On the other hand, very dim stars were less defined and lacked color, but were not as small as we might expect from looking at the brighter stars. This seems to support the notion that more than one detection threshold exists for the eye, and that *t = mag 4.9* reflects a threshold for the less sensitive but color-sensitive cells of the eye.**

**Our visual results gave us added confidence in the results of the model, added appreciation for Galileo's abilities at the eyepiece, and added understanding of why he interpreted the images he saw to be physical bodies of stars. In most cases what was seen visually very much matches simulated views created from the model, such as is seen in Figure 5. The main weakness of the model appears, not surprisingly, when approaching the model's magnitude limit, as seen in the case of ο Cyg.**

would likely conclude that each double star is physically separated by a value on order of $s_2$.

From Table 1 it seems clear that had the stars been studied more carefully during Galileo's time (or perhaps even if Galileo had simply published his observations of Mizar, the Trapezium, and other multiple star groupings), the concept of physically grouped double and multiple star systems, composed of stars of varying sizes, would have necessarily arisen. Under Galileo's interpretation of the images of stars seen through his telescopes being the physical globes of those stars, observations of double stars in 1617 would have led to the conclusion that double stars were closely spaced compared to sun-planet distances, let alone sun-star distances (Figure 7). Granted, that conclusion would have been a distorted one in which the stars were both too close to Earth and too close to each other; nevertheless, it is no less distorted than Galileo's assumption that all stars are single and identical to the sun. As it was, the idea of double star systems did not actually arise until the time of William Herschel, nearly two centuries after Galileo made his 1617 observations (Hirshfeld 2001, pp. 187-189).[5]

## 6. POSSIBLE CONCLUSION: THE FIRMAMENT EXISTS

There is also the possibility that more careful study of the stars may have simply undermined Galileo's ideas about the structure of the universe. In light of the likely data a search for double stars would have gathered, an argument could have been made in the 17th century that the concept of physically grouped star systems was “ad hoc”. After all, the physical components of a star like Mizar, as determined using Galileo's methods, were very close. But more importantly, Galileo had observed the Trapezium.

---

5 This is the prevailing view. See comments about recent research in the Acknowledgments at the end of the paper.

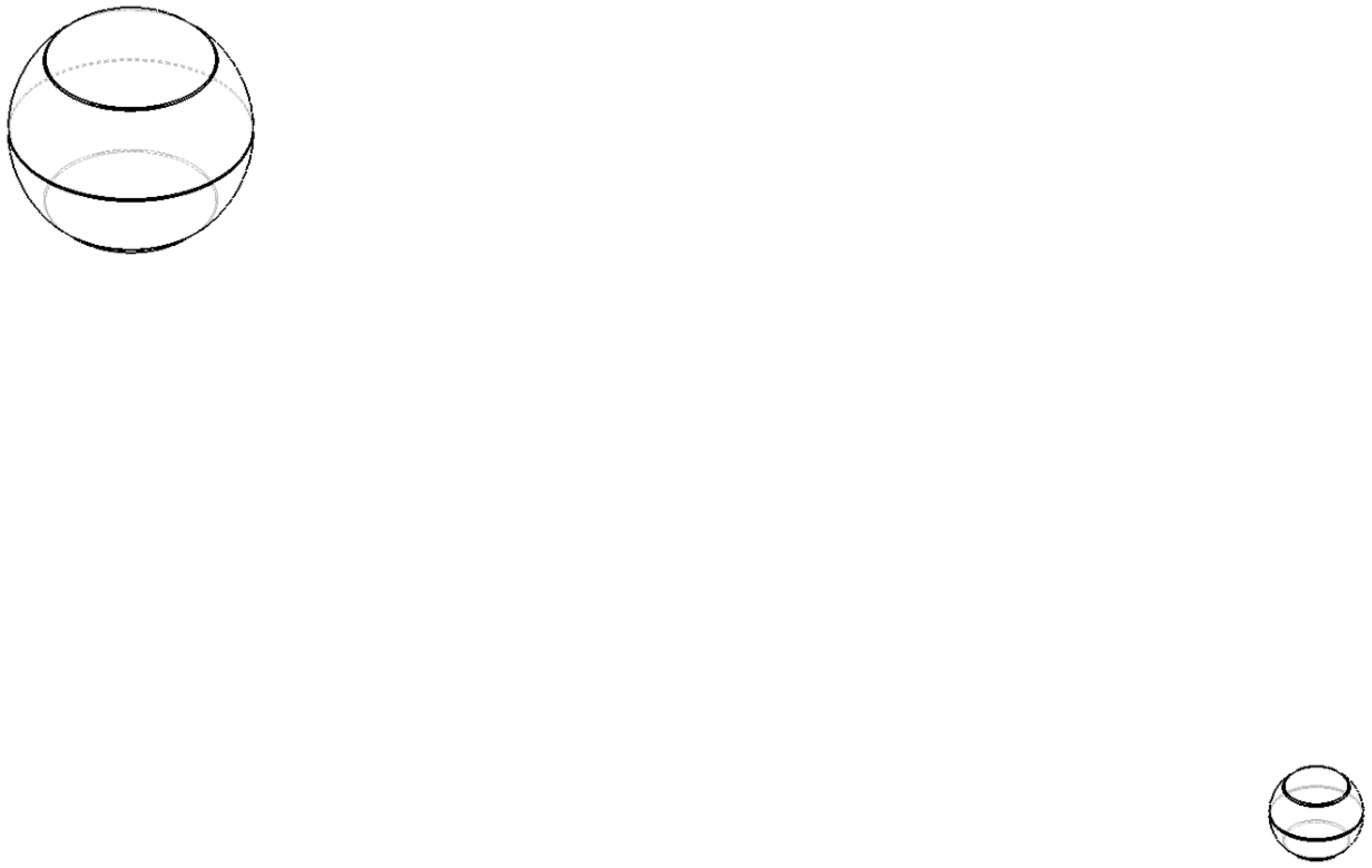

**FIGURE 7: To convey a sense of scale, here we provide a "three-dimensional" diagram of the ß Cyg double star system based on Galileo's ideas about stellar sizes and distances and the data derived from our model's output. The system is shown for the case of $s_1 = 0$, so this shows the stars in their closest possible configuration. Earth lies in the plane of the equators of the globes in the figure, and along a line perpendicular to the stars' line of separation. The larger star is the size of the sun.**

As mentioned earlier, Galileo precisely mapped the stars of the Trapezium along with other nearby stars, and while he did not record absolute diameter measurements of the stars in the Trapezium region, he did note that their diameters varied by a factor of four or five (*Opere*, III, Pt. 2, p. 880). Under Galileo's assumption that the Trapezium stars are the same size as the sun, if their sizes vary by a factor of four or five then their distances must vary by a factor of four or five. This means that even if the stars are distant in the extreme their parallactic changes would be larger than their separations and easily detectable.[6] If the Earth is moving, the lack of parallax in the Trapezium stars means they must be a physically grouped system. But if the Trapezium stars are a physically grouped multiple star system, their physical diameters must actually vary by a factor of four or five. Such a size variation is a significant break from Galileo's ideas.

In fact, it is a fatal break. Everything Galileo has to say about stars is constructed upon the assumption that stars are essentially the same size as the sun. The more stars can vary in size, the shakier the entire system of thought becomes. In the *Dialogue* Galileo implies only a factor of six variation in size between first and sixth magnitude stars (p. 359) -- if stars can vary in size by a factor of four or five then Galileo's distances calculated from the assumption that stars are suns are meaningless.

This problem would not be limited to the Trapezium. Even a star from Table 1 such as β Cyg would appear to Galileo to have significant size variation, and that table is limited to those whose fainter companions are brighter than magnitude 4.9. Galileo could observe and precisely record the position of an object as faint as Neptune, while it was among objects as bright as the members of the Jovian system (Standish and Nobili 1997).

6 Under Galileo's assumption that all stars are the size of the sun, even if the faintest star is as distant as 40,000 AU, and the brightest is only four times closer, we still find $\Delta > 15''$.

Any search for double stars would doubtlessly find bright stars with fainter companions, presenting more problems for Galileo's ideas.

The conclusion astronomers in the 17th century might well reach in light of all this is that Galileo's ideas regarding the structure of the universe are fundamentally flawed and the universe does not, in fact, consist of sun-like stars distributed throughout space. Perhaps a more persuasive interpretation of the data might be that they support Copernicus' original description of the stars being arranged on a sun-centered firmament. Another persuasive interpretation might be that the Earth is not moving, and that a geocentric theory compatible with telescopic discoveries, such as that of Tycho Brahe, is the best explanation of the data (Graney 2008b).

## 7. EARLY OBSERVATIONS OF THE STARS: WORTH FURTHER STUDY

Along with these possible conclusions we should also consider that other lines of evidence could be gleaned from observations of the stars. One of us (Graney 2008a) has recently argued in this journal that Galileo's ideas about the stars can be used to calculate how the number of stars visible to the naked eye increase with magnitude, and that the results of such calculations appear to bolster Copernican ideas. Moreover, Galileo was not the only astronomer of his day to observe that stars showed distinct disks as seen through the telescope, to interpret those disks as being the physical bodies of stars, and to use those observations to draw conclusions about the structure of the universe. During the same time period the German astronomer Simon Marius also observed that stars showed noticeable disks, with bright stars showing them most clearly; like Galileo he interpreted those disks as being the physical bodes of stars; and like Galileo he also argued that the disk-like appearance of stars gave indication of their distance (Marius 1614, pp. 46-49).[7]

It seems clear that observations of stars in the early 17th could lead to many different conclusions concerning the stars. Further investigation of this area may yield insight into what Galileo was thinking and how he came to support the Copernican theory, as well as insight into what his colleagues were thinking.

## 8. CONCLUSION

We conclude that greater telescopic study of the stars at the dawn of telescopic astronomy -- or even simply the publication of Galileo's observations of Mizar, the Trapezium, and other double and multiple star groupings -- would have had a significant impact on the development of conceptions regarding stars and the structure of the universe. The issue of the lack of parallax among close doubles whose distances had been determined incorrectly (because Galileo interpreted the images

7 However, literature on Marius's stellar observations is even more sparse than literature on Galileo's stellar observations. See Dreyer 1909, p. 191 for a rare mention of Marius's stellar observations.

of stars seen through his telescopes as being the physical globes of those stars) might well have had a significant impact on conclusions reached regarding the stars. This in turn would have led astronomers such as Galileo to abandon the idea that stars are essentially single bodies identical to the sun and, instead, to view stars as being generally sun-like while varying in size and occasionally being closely paired. That would have been a step in the right direction. But it is also possible that greater telescopic study of stars during Galileo's time may have led to steps very much in the wrong direction -- away from the idea that stars were suns at all, and possibly even away from the idea of a moving Earth. The stars could be a fruitful target of study, even in the early 17$^{th}$ century, and this neglected aspect of astronomy's history is worth further study in order to help determine how modern ideas concerning the structure of the universe of stars came to exist.

## ACKNOWLEDGMENTS

We would like to thank the U.S. Naval Observatory, and in particular Brian Mason, for assistance with the selection of double stars meeting our criteria, and for providing other data on those stars where needed. We would also like to thank Jefferson Community & Technical College and Otter Creek Park, both of Louisville, Kentucky (USA) for their support of Otter Creek Observatory, and in particular the provost of Jefferson, Dr. Diane Calhoun-French, and the Jefferson Technology and Related Sciences (TARS) Division, both of whom have provided funds for the equipment used in this research.

Lastly, we would like to thank an anonymous referee for providing an update on recent research concerning when the idea of double star systems first arose:  While the usual view has been that the idea arose with Herschel (see footnote 5), very recent research indicates that a number of astronomers in the first half of the 17$^{th}$ century were interpreting telescopic discoveries of double stars in terms of star systems, and that Athanasius Kircher made use of double star findings in order to represent a non-Copernican cosmos in his writings.  See Siebert 2006.